# The Effect of Dust and Sand on the 5G Terrestrial Links

Esmail M M Abuhdima
Comp. Sc., Phys., and Engineering
Benedict College
Columbia, USA
esmail.abuhdima@benedict.edu

Gurcan Comert
Comp. Sc., Phys., and Engineering
Benedict College
Columbia, USA
gurcan.comert@benedict.edu

Pierluigi Pisu
Int. Center for Automotive Research
Clemson University
Greenville, USA
pisup@clemson.edu

Chin-Tser Huang
Computer Science and Engineering
University of South Carolina
Columbia, USA
huangct@cse.sc.edu

Ahmed El Qaouaq
Comp. Sc., Phys., and Engineering
Benedict College
Columbia, USA
ahmed.el-qaouaq34@my.benedict.edu

Chunheng Zhao
Int. Center for Automotive Research
Clemson University
Greenville, USA
chunhez@clemson.edu

Shakendra Alston
Comp. Sc., Phys., and Engineering
Benedict College
Columbia, USA
Shakendra.Alston68@my.benedict.edu

Kirk Ambrose
Comp. Sc., Phys., and Engineering
Benedict College
Columbia, USA
Kirk.Ambrose61@my.benedict.edu

Jian Liu
Computer Science and Engineering
University of South Carolina
Columbia, USA
jianl@email.sc.edu

*Abstract*—Wireless connections are a communication channel used to support different applications in our life such as microwave connections, mobile cellular networks, and intelligent transportation systems. The wireless communication channels are affected by different weather factors such as rain, snow, fog, dust, and sand. This effect is more evident in the high frequencies of the millimeter-wave (mm-wave) band. Recently, the 5G opened the door to support different applications with high speed and good quality. A recent study investigates the effect of rain and snow on the 5G communication channel to reduce the challenge of using high millimeter-wave frequencies. This research investigates the impact of dust and sand on the communication channel of 5G mini links using Mie scattering model to estimate the propagating wave's attenuation by computing the free space loss of a dusty region. Also, the cross-polarization of the propagating wave with dust and sand is taken into account at different distances of the propagating length. Two kinds of mini links, ML-6363, and ML-6352, are considered to demonstrate the effect of dust and sand in these specific operating frequency bands. The 73.5 GHz (V-band) and (21.5GHz (K-band) are the ML-6352 and ML-6363 radio frequency, respectively. Also, signal depolarization is another important radio frequency transmission parameter that is considered heroin. The numerical and simulation results show that the 5G ML-6352 is more effect by dust and sand than ML6363. The 5G toolbox is used to build the communication system and simulate the effect of the dust and sand on the different frequency bands.

## I. INTRODUCTION

Various applications consider the 5G frequency band as the connected wireless channel. When it comes to vehicle-to-vehicle communication, we are sending signals using the 5G mm-wave communication channel. Recently, the radio frequency of the point-to-point connection is in the millimeter frequency band. The features of new communication technology require a high-speed transmit signal with good quality, so the 5G opens the door to support different applications in our life. The smaller wavelength allows mm-wave to be more affected by weather factors such as rain, snow, hail, dust, and sand [1]. The main focus of this research is studying the effect of dust and sand on the operation of 5G mini links communication systems. Previous work displayed the impact of dust and sand on Microwave connection and cellular mobile covered signals by calculating the attenuation factor in dB/km. The attenuation was calculated in terms of different parameters such as visibility, height, particle size, and dielectric constant [2], [3], and [4].

Another research considers the effect of dust and sand on the 5G wireless communication channel between the connected vehicles. This research concludes that the attenuation of the propagating signal increases when the operating frequency, the concentration of the dust, and the particle size of sand are increased. The 5G (28GHz) mm-wave communication channel is more affected by dust and sand storms than the DSRC (5.9GHz) channel [5].

This work considers the effect of cross-polarization with the attenuation factors that affect the propagating signal through a dusty region. The research aims to demonstrate the impact of dusty environments on wireless communication channels, especially those using a 5G frequency band. For this reason, a real measurement of the dielectric constant, particle size range, and concentration of dust of the desert area are required to compute the accurate attenuation factor for any specific region.

This research is significant because sand and dust storms are problems that occur worldwide. This research aims to determine the attenuation and cross-polarization discrimination (XPD) effect by sand and dust storms on the 5G Communication Channel of ML-6363 and ML-6352 links. The proposed path loss is used to simulate the impact of dust and sand during the degradation of the transmit power signal. Finally, the Mathlab 5G toolbox will be used to model the research problem.

This paper is separated into 4 –four main sections. Section II generally discusses the main topic of the research that studied the propagating mm-waves in dusty regions. The simulation and the numerical results are introduced in section III. Finally, the conclusion and recommendation are shown in Section IV.

## II. MILLIMETER WAVES AND DUSTY REGION

The degradation of the power of transmit signal is defined as an attenuation. In other words, the attenuation is known as the general loss in the power of transmitting signal. A general explanation of attenuation is that transmitting parameters such as visibility, frequency, wavelength, and permittivity can vary and cause a weakening of an mm-wave as it propagates through a dusty region. The primary source of the dust and sand particles on the planet is the deserts. In the U.S., the largest desert is nearly 190,000 square miles, while the entire continent of Australia has deserts covering 529,000 square miles [6]. In Northern Africa, the Sahara Desert is approximately 3.5 million square miles and is the world's largest desert. About twenty percent of the world's deserts are covered in the sand [7]. As mentioned previously, the attenuation is dependent upon the parameters stated. It is known that the amount of time that a wave passes a certain point in a specific amount of time is defined as a frequency, which is measured in hertz (Hz). However, humidity is the amount of atmospheric moisture present. According to collected data, the humidity changed from 0% (dry weather) to 100% as worst-case. The height ($h$), another critical factor, defines the distance from the ground to where the antenna is placed. Visibility is the distance that a human can see when dust and sand are present. The particle size of sand ranges from 0.0625–2 mm, and the particle size of dust is from 1-100μm [4]. The Equation (1) represents the visibility at a certain reference height $h_0$ and reference visibility $v_0$ as

$$V^\gamma = v_0^\gamma \left[\frac{h}{h_0}\right]^b \quad (1)$$

where $\gamma$ and b are a constant. The dielectric constant is another important measured value to examine the effect of dust and sand. The dielectric constant is based on the sand's amount of moisture. Sand and dust can affect the dielectric constant, and this dielectric constant varies when the amount of moisture in the sand/dust changes. As the permittivity of the sand or dust changes, the dielectric constant changes as well. The mineral and chemical composition of the dust and sand particles is used to compute the dielectric constant. Looyenga equation is used to calculate the complex permittivity $\varepsilon_m$ as

$$\varepsilon_m^{\frac{1}{3}} = \sum_{i=1}^n v_i \varepsilon_i^{\frac{1}{3}} \quad (2)$$

where $v_i$ and $\varepsilon_i$ are the relative volume and the complex dielectric constant of the $i^{th}$ sample. It can write the permittivity of the free space with dust and sand as

$$\varepsilon = \varepsilon' + j\varepsilon'' \quad (3)$$

where $\varepsilon'$ is the dielectric constant, and $\varepsilon''$ is the dielectric loss factor. The effect of the humidity on the permittivity of the dusty free space is written as [2]

$$\varepsilon_1 = \varepsilon' + 0.04H - 7.78\times10^{-4} H^2 + 5.56\times10^{-6} H^3 \quad (4)$$

$$\varepsilon_2 = \varepsilon'' + 0.02H - 3.71\times10^{-4} H^2 + 2.76\times10^{-6} H^3 \quad (5)$$

where $H$ is the air relative humidity in percentage. The general permittivity is defined as

$$\varepsilon = \varepsilon_1 + \varepsilon_2 \quad (6)$$

### A. Attenuation of propagating wave

The geometry of this research is shown in Figure 1. The terrestrial link has two antennas. The first one is in point A and the second one is in point B. It is considered that the operating frequencies belong to the 5G frequency band. It is regarded as two mini links that used different operating frequencies. The main goal of this paper is to demonstrate the effect of dust and sand on the operation of these 5G mini links. It is known that the radio path loss is an important value to estimate the radio coverage signal. In other words, the received signal's strength depends on the value of the channel path loss. For a clear line of sight (LOS), the range between point A (transmitter) and point B (receiver) in typical weather is limited by [8]

$$path\,loss_{(dB)} = 20\log\left[\frac{4\pi d}{\lambda}\right] \quad (7)$$

where $d$ is the range in meters and $\lambda$ is the wavelength in meters. In the case of dust and sand, adding dust and sand effect into the channel path loss is required. There are five methods that calculate the wave attenuation in the free space with dust and sand. These five formulas produce a similar result of signal attenuation, especially at frequencies above 30GHz and visibility less than 20 meters [9]. The particle radius and the particle size distribution are major important factors that have more effect on the wave attenuation. The Mie method is considered herein to investigate the effect of dust and sand because it takes the mutual interaction phenomenon into account. This method shows the attenuation and does so by considering factors including visibility, height, particle size, humidity, dielectric constant, and frequency. Each of these factors has equations to calculate them, but the Mie Model integrates them into one equation used to find the attenuation. The dust and sand attenuation is defined by

$$A_d = \frac{a_e f}{v}\left[C_1 + C_2\, a_e^2\, f^2 + C_3\, a_e^3\, f^3\right] \quad \frac{dB}{km} \quad (8)$$

where $a_e$ is the radius of the particle in meters, $v$ is the visibility in a kilometer, f is the operating frequency in GHz, $C = 2.3\times10^{-5}$, $\gamma = 1.07$, $g = 1.07$, $b = 0.28$,

$$C_1 = \frac{6\varepsilon_2}{(\varepsilon_1+2)^2 + \varepsilon_2^2}$$

$$C_2 = \varepsilon_2\left[\frac{6}{5}\frac{7\varepsilon_1^2 + 7\varepsilon_2^2 + 4\varepsilon_1 - 20}{\left[(\varepsilon_1+2)^2+\varepsilon_2^2\right]^2} + \frac{1}{15} + \frac{5}{3\left[(2\varepsilon_1+3)^2+4\varepsilon_2^2\right]}\right]$$

and $C_3 = \frac{4}{3}\left[\frac{(\varepsilon_1-1)^2(\varepsilon_1+2)+\left[2(\varepsilon_1-1)(\varepsilon_1+2)-9\right]+\varepsilon_2^4}{\left[(\varepsilon_1+2)^2+\varepsilon_2^2\right]^2}\right].$

The attenuation of dust and sand can be written as

$$\alpha_{(dB)} = \int_0^d A_d\, dl \quad (9)$$

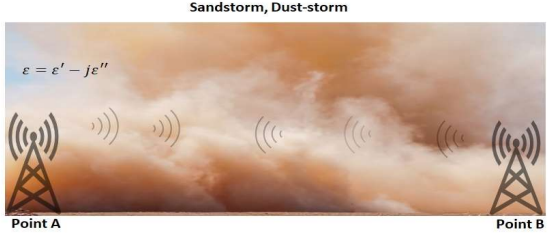

Fig. 1. 5G terrestrial link

The proposed model that is used to calculate the free space path loss with dust and sand can write as

$$path\,loss_{(dB)} = 20\log\left[\frac{4\pi d}{\lambda}\right] + \alpha \quad (10)$$

The Equation (11) represents the free space path loss of the 5G communication channel. This path loss is frequency, distance, dust, and sand dependent, and it increases with distance and visibility decreases. Also, the path loss increases when the radius of particular increases.

$$Path\,Loss_{(dB)} = 92.44 + 20\log(f) + 20\log(d) + \alpha \quad (11)$$

where $f$ is the operating frequency in GHz, $d$ is the distance between transmitter and receiver in km, and $\alpha$ is the attenuation of dust and sand in $dB$. It is found that if the visibility is large (normal weather), the value $\alpha$ is negligible.

*B. Cross polarization*

Polarization of transmitting signal refers to the direction that a wave travels. Co-polarization is the direction in which the wave is wanted to travel or the desired direction while cross-polarization is perpendicular to the desired direction. For example, if the desired polarization is horizontal, then the cross-polarization of that would be vertical. There is an intended polarization that most antennas have. Cross polarization is measured in negative dB to show how many decibels the power level is below the desired polarization. Cross polarization is important to minimize the interference of the waves. Cross-polarization discrimination is defined by [10]

$$XPD = 10\log_{10}\left|\frac{1 + 2m\cos\phi + m^2}{1 - 2m\cos\phi + m^2}\right| \quad (12)$$

where $m$ and $\phi$ defined as

$$m = e^{-|\alpha_h - \alpha_v|d} \quad (13)$$

$$\phi = (\phi_h - \phi_v)d \quad (14)$$

where $\alpha_v$ is the attenuation of vertical polarization, $\alpha_h$ is the attenuation of horizontal polarization, $\phi_v$ is the phase constant of vertical polarization, $\phi_h$ is the constant of horizontal polarization and $d$ is the propagating path between transmitter and receiver.

## III. SIMULATION AND NUMERICAL RESULT

It is known that the dielectric constant is an important parameter to estimate the effect of dust and sand. In other words, the complex permittivity $(\varepsilon)$ is used to compute the attenuation factor and the cross-polarization during dust and sand storms. It is considered to install these 5G mini links in the studied region located in North Africa. The real average density and the real complex permittivity of the studied region are $2.5764\ g/m^3$ and $6.3485 - j0.0929$ respectively. The average and maximum sizes of particles of different samples are shown in Table I.

TABLE I. Maximum size particles

| Sample No | $r_{avg}$ | $r_{max}$ |
|---|---|---|
| 1 | 94.43 $\mu m$ | 538.04 $\mu m$ |
| 2 | 64.34 $\mu m$ | 159.61 $\mu m$ |
| 3 | 25.23 $\mu m$ | 128.68 $\mu m$ |

Table I shows that the maximum radius of a sand particle is $538.04\ \mu m$. The technical transmission parameters of both kinds of 5G mini links are present in the Table II.

TABLE II. Parameters of 5G terrestrial links

| Descriptions | K-Band 1.8 km | V-Band 1.8 km |
|---|---|---|
| Transceiver name and manufactured | Mini Link ML-6363, Ericsson | Mini Link ML-6352, Ericsson |
| Frequency | 21.8 GHz | 73.5 GHz |
| Base Capacity | 28 Mbps | 100 Mbps |
| Antena type | Directional ANT3 0.6 23 HP | Directional ANT2 A 0.3 80 HP |
| Polarization | Vertical | Vertical |
| Antenna Gain | 40.7 dBi | 46.5 dBi |
| Max Tx power | 20 dBm | 15 dBm |
| 10^-6 BER Received threshold | -79dBm | -75dBm |

*A. Attenuation of dust and sand*

The Equation (11) was used to simulate the effect of dust and sand on the path loss of the 5G Mini links. The effect of dust and sand on the ML 6363 (21.8GHz) is evident when the visibility is less than 80 meters. This effect increases sharply when the visibility value is less than 20 meters, as shown in Fig. 2. For the ML 6352 (73.5 GHz), the strong effect of dust and sand accrued at the visibility of fewer than 30 meters, as shown in Fig. 3. It is shown that the impact of dust and sand is similar on both mini links when the visibility is 20 m and above, but when the visibility is less than 20m, the ML6352 is more effective in dusty storms. Also, the operation of these mini links is tested when the radius of sand particles changes, as shown in Figs. 4 and 5. The path loss of the communication channel increased when the radius of the particle increased. In dry weather (0% humidity), The path loss measures a lower quantity in comparison with 60% and 100 % humidity.

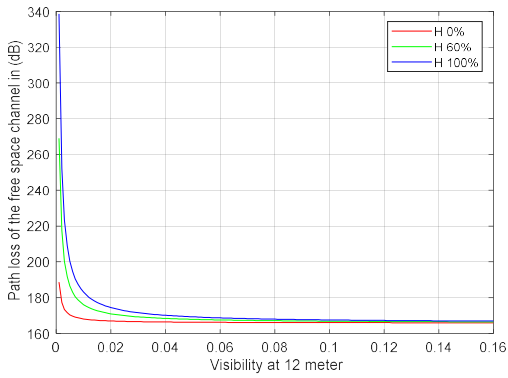

Fig. 2. Path loss of channel with visibility

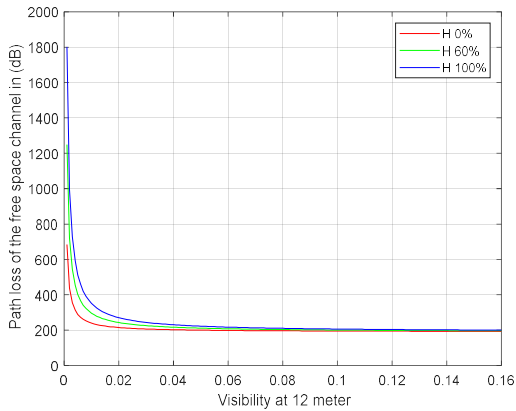

Fig. 3. Path loss of channel with visibility

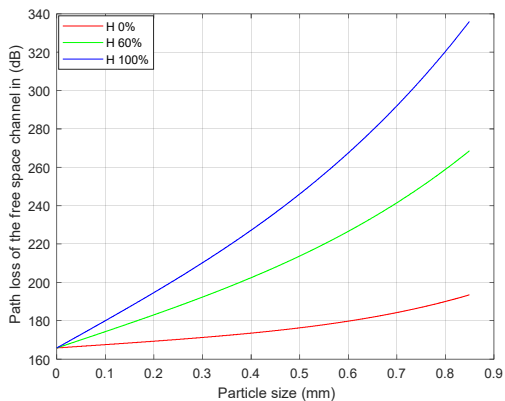

Fig. 4. Path loss of ML 6363 (21.8GHz)

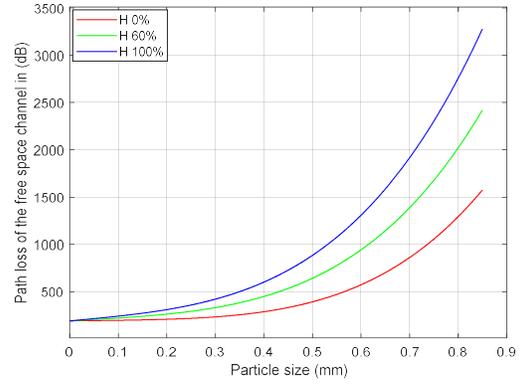

Fig. 5. Path loss of ML 6352 (73.5GHz)

### B. Cross polarization

Also, the effect of cross-polarization on the operation of both links (ML 6363 and ML 6352) is demonstrated by using (12), (13) and (14). In this simulation, it is considered that the complex permittivity of the dry dust and 4 percent water is $\varepsilon = 5.23 - j0.26$ and $\varepsilon = 6.23 - j0.57$ respectively [10]. The propagating length path $(d)$ is considered 1.8, 5, and 20km. The simulation result for ML6363 and ML6352 is shown in Figs. 6 and 7, respectively. The solid line for dry weather and dotted line for 4% water weather. For ML6363, the effect of sandy weather with visibility less than $12\,m$, $32m$ and $131m$ seriously effects $1.8km$, $5km$ and $20km$ propagating length respectively. A change in dust humidity from 0% to 4% produces a decrease in the cross-polarization discrimination by $3\,dB$ as shown in Fig. 6. Similarly, the effect of sandy weather with visibility less than $39\,m$, $106m$ and $406m$ seriously affects the ML6352 for $1.8km$, $5km\ and\ 20km$ propagating length, respectively. A change in dust humidity from 0% to 4% has produced a decrease in the cross-polarization discrimination by $2.5\,dB$ as

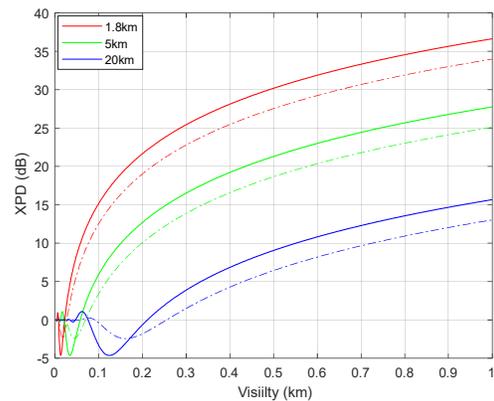

Fig. 6. Cross polarization of circularly polarized wave at 21.8 GHz.

shown in Fig. 7

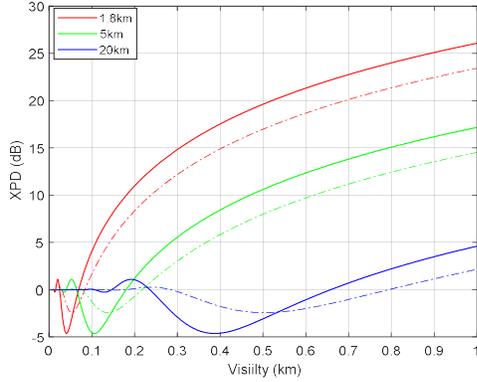

Fig. 7. Cross polarization of circularly polarized wave at 73.5 GHz.

*C. Simulation Study*

MATLAB and Simulink are used to model the 5G mini links as the transmitter, receiver, and communication channel. Point A and point B in Fig. 1 are named transmitter and receiver respectively in the Simulink model, as shown in Fig 8. This figure shows the model of the point-to-point 5G communication link. The simulation has been made using Simulink on MATLAB, and it consists of different blocks models. The three main parts of this Simulink connection are the transmitter, receiver, and communication channel. The transmitter block consists of a binary generator source, modulator, high power amplifier, and antenna to transmit the signal through free space. The communication channel consists of the path loss equation (11), implemented in the system as a function block where the signal has input from the transmitter. It is modified based on different parameters such as frequency, distance, visibility, and particle size. The last part of the system is the receiver, where the signal is being transmitted through the channel and received by the antenna block to deliver to the demodulator.

Fig. 9 shows the spectrum analyzer screen where the transmit signal is in the blue line, and the received signal is in the red line. It is configured that the transmit channel power is $30\,dBm$ and the antenna gain $40.7\,dB$. The parameters of ML6363 that are shown in TABLE II are considered. It is selected that $f = 21.8\,GHz, d = 1.8\,km, V = 10\,m$, and the humidity is $60\%$. It is seen that the received channel power is $691\,dBm$ and $BER = 4 \times 10^{-5}$, so the total loss is $661\,dBm$. When the visibility sets to $0.1\,m$, the receiver lost the signal and $BER = 2.4 \times 10^{-1}$, but the received channel power is measured to $1496\,dBm$ at dry weather and the $BER = 5.7 \times 10^{-5}$. In the case of ML6352 (73.5GHz), It is found that the received channel power is $1099\,dBm$, and the $BER = 5 \times 10^{-5}$ when $f = 73.5\,GHz, d = 1.8\,km, V = 10\,m$ and the humidity is $60\%$. When the visibility sets to $0.1\,m$, the receiver lost the signal and the $BER = 4.998 \times 10^{-1}$ in both cases, $60\%$ and $0\%$ humidity. In general, both 5G mini links are affected by the sandy weather, especially when the visibility is less than $10\,m$. Also, the ML-6352 is more effect by dust and sand storms than ML-6352.

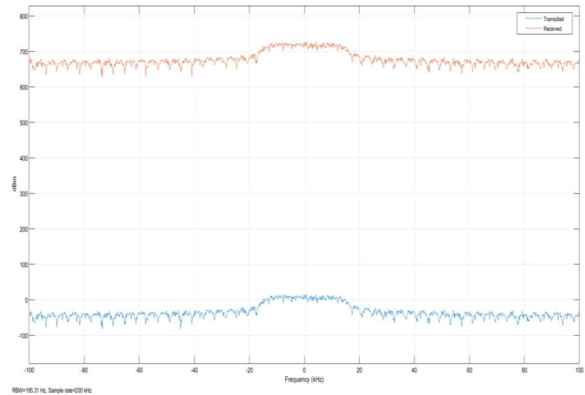

Fig. 9. Transmitted and received signal of ML6363 (21.8 GHz).

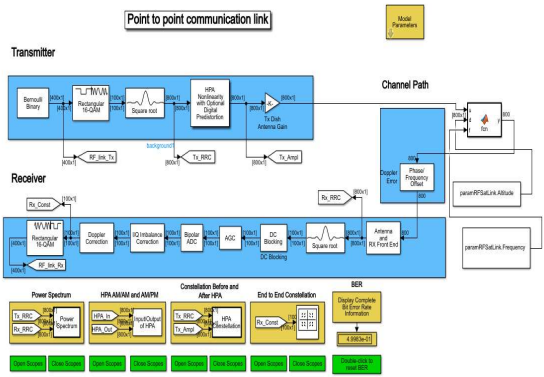

Fig. 8. Point to point communication link.

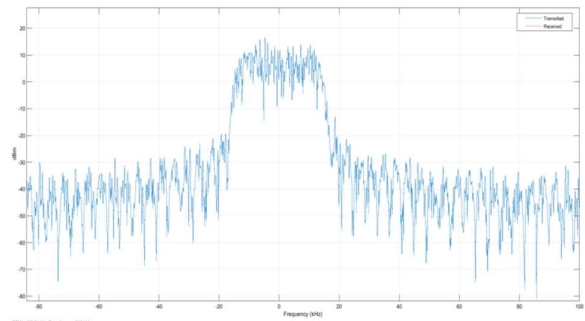

Fig. 10. Transmitted and received signal of ML6352 (73.5 GHz).

## IV. Conclusion

In this research, the effect of dust and sand on the operation of the 5G mini links is investigated. The effects of dusty storms on the radio frequency of the ML6363 and ML6352 are demonstrated in terms of length of propagation wave, visibility, particle size, and operating frequency. The numerical and simulation result shows that the path loss of the communication channel increase when the frequency increases, distance increases, and visibility decreases. It is seen that the ML6363 is seriously affected by the dust and sand when the visibility is less than 12m at a distance of 1.8km. The ML6352 is strongly affected by dust and sand when the visibility is less than 39m at a distance of 1.8km. This research recommended computing the real value of attenuation for different desert regions of the united states. This value will help design the wireless connection system to avoid any discount or loss of data.

## References

[1] A. M. Al-Saman, M. Cheffena, M. Mohamed, M. H. Azmi and Y. Ai, "Statistical Analysis of Rain at Millimeter Waves in Tropical Area," in IEEE Access, vol. 8, pp. 51044-51061, 2020, doi: 10.1109/ACCESS.2020.2979683.

[2] E. M. Abuhdima and I. M. Saleh, "Effect of sand and dust storms on GSM coverage signal in southern Libya," 2010 International Conference on Electronic Devices, Systems and Applications, 2010, pp. 264-268, doi: 10.1109/ICEDSA.2010.5503063..

[3] E. M. Abuhdima and I. M. Saleh, "Effect of sand and dust storms on microwave propagation signals in southern Libya," Melecon 2010 - 2010 15th IEEE Mediterranean Electrotechnical Conference, Valletta, 2010, pp. 695-698.

[4] I. M. Saleh, H. M. Abufares and H. M. Snousi, "Estimation of wave attenuation due to dust and sand storms in southern Libya using Mie model," WAMICON 2012 IEEE Wireless & Microwave Technology Conference, Cocoa Beach, FL, 2012, pp. 1-5.

[5] E. Abuhdima, Gurcan Comert, Ahmed Elqaouaq, Shakendra Alston, "Impact of Weather Conditions on 5G Communication Channel under Connected Vehicles Framework," in press.

[6] Britannica, T. Editors of Encyclopaedia (2014, August 27). List of deserts. Encyclopedia Britannica. https://www.britannica.com/topic/list-of-deserts-1854209

[7] Britannica, T. Editors of Encyclopaedia (2019, December 9). Great Basin. Encyclopedia Britannica. https://www.britannica.com/place/Great-Basin

[8] Mohammed B. Majed, Tharek A. Rahman and Omar Abdul Aziz, "Propagation Path Loss Modeling and Outdoor Coverage Measurements Review in Millimeter Wave Bands for 5G Cellular Communications," International Journal of Electrical and Computer Engineering (IJECE), Vol. 8, No. 4, August 2018, pp. 2254~2260.

[9] Abdulwaheed Musa*, Saad O. Bashir, and Aisha H. Abdalla, "Review and Assessment of Electromagnetic Wave Propagation in Sand and Dust Storms at Microwave and Millimeter Wave Bands — Part II," Progress In Electromagnetics Research M, Vol. 40, 101–110, 2014.

[10] S. Ghobrial and S. Sharief, "Microwave attenuation and cross polarization in dust storms," in IEEE Transactions on Antennas and Propagation, vol. 35, no. 4, pp. 418-425, April 1987, doi: 10.1109/TAP.1987.1144120.